\documentclass[12pt]{iopart}

\pdfoutput=1

\usepackage{graphicx}
\usepackage{amsfonts}
\usepackage{amssymb}

\begin{document}
\title{Pathology in WKB wave function for tunneling assisted by gravity}
\author{Stefano Ansoldi$^{1,2,3}$ and Takahiro Tanaka$^{4,5}$} 
\address{$^{1}$Department of Mathematics, Computer Science, and Physics, University of Udine, I-33100 Udine, Italy}
\address{$^{2}$Istituto Nazionale di Fisica Nucleare (INFN), Sezione di Trieste, Italy}
\address{$^{3}$Institute for Fundamental Physics of the Universe (IFPU), Trieste, Italy}
\address{$^{4}$Department of Physics, Kyoto University, Kyoto 606-8502, Japan}
\address{$^{5}$Center for Gravitational Physics and Quantum Information, Yukawa Institute for Theoretical Physics, Kyoto University, Kyoto 606-8502, Japan}
\date{\today}

\begin{abstract}
There are several exotic tunneling processes that can be realized only by incorporating the effect of gravity. 
Here, we point out that we encounter difficulties in constructing the WKB wave function, 
once we try to describe quantum fluctuations around the semi-classical tunneling path. 
We present examples of pathology in the true vacuum decay/upward tunneling, the false vacuum decay catalyzed by a black hole and the tunneling with black hole/wormhole production. 
\end{abstract}

\submitto{\CQG}

\maketitle

\section{Introduction}

Tunneling is, perhaps, one of the most surprising processes in quantum physics~\cite{bib:doi:10.1063/1.1510281}, with applications that range from the study of the transitions between string vacua in the landscape~\cite{bib:Ahlqvist:2010ki} to the principles of operation of the scanning tunneling microscope~\cite{bib:BinnigNobel}. A concrete understanding of tunneling would be obtained by explicitly constructing the wave function under the potential barrier where classical motion is forbidden. 
In this paper we consider the validity of the WKB approximation in this construction of the wave function along the dominant escape path that minimizes the Euclidean action 
in tunneling processes. 
Even when considering systems with many degrees of freedom, we can 
identify a collective coordinate corresponding to the direction along the dominant escape path. 
The wave function for this collective coordinate is given by the exponential of the negative of the classical Euclidean action. 
Then, at the lowest order in the quantum corrections, one can evaluate the ratio in the squared amplitudes of the wave function between the initial and final configurations corresponding to the both ends of the dominant escape path, which will tell us about the tunneling probability.   
At this level there seems to be no obstacle in constructing the WKB wave function and no doubt about its interpretation. 
Then, we turn to the remaining degrees of freedom and 
try to determine the wave function away from the dominant escape path. 
The shape of the wave function in the direction perpendicular to the dominant escape path also has physical relevance because it determines the quantum state after tunneling. 
Furthermore, the one-loop correction to the tunneling rate is imprinted in the wave function at this order. 

When we consider a field theoretical tunneling on a fixed Minkowski background, 
the procedure to construct the WKB wave function including quantum fluctuation around the dominant escape path is well-understood~\cite{Banks:1973ps,Banks:1973uca,Gervais:1977nv,Bitar:1978vx,bib:PhReD1994..49..1039Y}.
However, when we consider the tunneling with gravitational degrees of freedom~\cite{bib:PhReD1994..50..6444S}, 
which will be our focus in this work, there seem to exist some cases in which it is less clear 
whether or not one can construct a fully justified WKB wave function.  
In the standard false vacuum decay with gravity the effect of the geometrical degrees of freedom on the tunneling process can be considered as a minor correction, and it is, then, not surprising that the standard approach to construct the WKB wave function can be 
extended consistently~\cite{bib:PhReD1994..50..6444S}. However, it is known since a long time that 
there exist some tunneling processes where some non-trivial Euclidean geometry plays 
an important role in constructing the Euclidean solution corresponding to the dominant escape path (i.e., the instanton). 
Examples categorized in this class are the true vacuum decay/upward tunneling~\cite{bib:PhReD1987..36..1088W}, the false vacuum decay catalyzed by a black hole~\cite{Gregory:2013hja}, and the tunneling with black hole/wormhole 
production~\cite{bib:NuPhy1990B339...417G}. 
Here, we raise a question whether or not the standard method to construct a WKB wave function 
can still be valid even in such cases. 

After recapitulating the general framework, we will concentrate on a few specific unconventional processes mentioned above. 
All processes have been extensively studied in literature, as they can have relevant phenomenological implications to understand the history of the early universe, inflationary models in particular. 
The upward tunneling plays an important role in the eternal inflation scenario~\cite{Garriga:2005av}, and also 
bears close relation with the Schwinger process in de~Sitter space~\cite{Garriga:1993fh,Frob:2014zka}, while  
the tunneling with black hole/wormhole 
production is connected to the process of baby/child-universe production. 

Keeping these ideas in mind, the paper is organized as follows. 
In section~\ref{sec:WKB} we briefly discuss how to construct the WKB wave function for systems with multiple degrees of freedom. We exemplify the process using the simplest system with two degrees of freedom, where one of them is the tunneling degree of freedom, and the other is a spectator one. Then, we just give the generalization of the final result to the case of field theoretical tunneling with gravity~\cite{bib:PhReD1994..49..1039Y,bib:PhReD1994..50..6444S}. 
Then, in section~\ref{sec:rev} we set the relevant technical background and notation by a quick review of the standard false vacuum decay through the spherical symmetric bubble formation. 
Then, in section~\ref{sec:wkbbre} we discuss a very simple example of upward tunneling which 
transforms a lower energy de~Sitter vacuum into another higher energy de~Sitter vacuum. 
In this process we find that 
a flip of sign of the lapse function is inevitable along the dominant escape path. 
Using this model, we explain the connection between the flip of the sign of the lapse function and the breakdown of the WKB approximation. 
Using this result, in section~\ref{sec:othpat}, we consider more phenomenologically relevant processes. 
We end with a summary and a concise discussion in section~\ref{sec:discon}.

\section{WKB wave function}
\label{sec:WKB}
In this section, we show how to obtain the WKB wave function in a system with multi-degrees of freedom. Not to unnecessarily complicate the discussion, we will consider a system composed of only two degrees of freedom, $\sigma$ and $\phi$, where $\sigma$ is to be regarded as the tunneling degree of freedom.
 Let the Hamiltonian be
\[
 H=\frac12\pi_\sigma^2+U(\sigma)+\frac12\left(\pi_\phi^2+m^2(\sigma)\phi^2\right)\,,
\]
where $\pi_\sigma$ and $\pi_\phi$ are the conjugate momenta of 
$\sigma$ and $\phi$, respectively. The potential $U(\sigma)$ 
is assumed to have a meta-stable minimum at $\sigma = 0$ (false vacuum), and we consider the tunneling of $\sigma$ away from this minimum. Let us consider an eigenfunction $|\Psi\rangle$ that satisfies 
\begin{equation}
    H|\Psi\rangle = E|\Psi\rangle\,,
\label{eq:wavequ}
\end{equation}
and focus, first, on the tunneling degree of freedom $\sigma$, neglecting $\phi$. 
Then, the unnormalized WKB wave function is obtained starting with the ansatz 
\begin{equation}
  \Psi_0(\sigma):=\langle \sigma|\Psi\rangle =\frac1{\sqrt{\pi_\sigma(\sigma)}}\exp\left[-\int^\sigma \pi_\sigma(\sigma') d\sigma'\right]\,.
  \label{eq:Psi0}
\end{equation}
Substituting the above expression into~(\ref{eq:wavequ}), we have 
\begin{equation}
    \pi_\sigma(\sigma)=\sqrt{2(U(\sigma)-E_0)}\,,  
    \label{eq:HJ}
\end{equation}
where $E_0:=U(0)$. 
By identifying $\pi_\sigma(\sigma)$ with $d\sigma/d\tau$, 
and by taking the derivative of the square of Eq.~(\ref{eq:HJ}) with respect to $\sigma$, we obtain the Euclidean equation of motion as
\[
    \frac{d^2\!\sigma}{d\tau^2}=\frac{dU(\sigma)}{d\sigma}\,. 
\]
Then, the solution 
\[
    \tau=\int^\sigma \frac{d\sigma'}{\sqrt{2 ( U(\sigma')-E_0 )}}\,, 
\]
gives a one-to-one correspondence between $\sigma$ and $\tau$. 
This Euclidean time coordinate fully parameterizes configurations along the dominant escape path of the WKB wave function.

Let us now consider the effect on the wave function produced by introducing the second degree of freedom $\phi$. 
For the tunneling from an initial meta-stable vacuum state, 
we can assume that the wave function is given in the form 
\[
    \Psi=\Psi_0(\sigma) \Phi(\phi,\tau(\sigma))\,, 
\]
with the standard Gaussian ansatz
\begin{equation}
  \Phi(\phi,\tau):=\left\langle\sigma(\tau),\phi|\Psi\right\rangle=
   \frac{e^{\delta E\tau}}{\sqrt{K(\tau)}}
      \exp\left(-\frac12 \frac{d\log K}{\, d\tau} \phi^2\right)\,,
\label{eq:Gauans}
\end{equation}
where $\delta E:=E-E_0$.
Substituting this expression into~(\ref{eq:HJ}), we have 
\[
     \left[\frac{d^2}{d\tau^2}+m^2\left(\sigma(\tau)\right)\right]K =0\,, 
\]
which is nothing but the Euclidean (linear) equation of motion for $\phi$. 
If we require the wave function for $\phi$ to be connected to the ground state 
near the false vacuum minimum at $\sigma=0$ and $\tau=-\infty$, we need to impose the boundary condition
\[
 \frac{d\log K}{d\tau} \to m(0)\,, \qquad \mbox{for}\quad \tau\to -\infty\,.
\]
Of course, the overall normalization of $K$ is irrelevant. Therefore, this condition is enough to completely determine the boundary condition for $K$. 
Under, the requirement that the wave function remains finite in the limit $\tau\to -\infty$, we find that we need to set $\delta E=m$. 

A crucial point for the analysis presented later in this paper, is that if either $K$ or $dK/d\tau$ vanishes for some value of $\tau$, then $d\log K/d\tau$ flips sign at this point, and, once $d\log K/d\tau$ becomes negative, the wave function 
becomes unnormalizable in the $\phi$-direction (please, see Eq.~(\ref{eq:Gauans})). Therefore, the WKB wave function cannot 
be trusted as a good approximation any more. Additionally, we note that, as long as $m^2(\sigma)$ is non-negative, the sign flip of $d\log K/d\tau$ is guaranteed not to happen. 
Here, we are implicitly assuming that 
the wave function along the path on which the Euclidean time $\tau$ 
monotonically increases. 
Indeed, the above Euclidean mode function $K$ is the one obtained by the analytic continuation, $t=-i\tau$, of the ordinary Lorentzian \emph{negative} frequency function. 

We now wish to generalize the above situation to the gravitational case. 
In the context of field theory, taking into account gravity in the above discussion becomes somewhat
more involved, but the final result is similar. 
Since what we now solve is the Wheeler-De~Witt equation, there is no `tunable' energy eigenvalue, {\it i.e.}, we do not have a term corresponding to $\exp (\delta E \, \tau)$ in Eq.~(\ref{eq:Gauans}). 
Instead, the spacetime curvature balances with the matter-energy contribution. 
The function $K$ of the example considered above is now promoted to a set of orthogonal 
Euclidean negative frequency mode functions, labeled by the generic index $i$ in what follows. 
In complete analogy with~(\ref{eq:HJ}), the wave function for a field $\phi$, which corresponds to some gauge invariant degrees of freedom, is given by~\cite{bib:PhReD1994..50..6444S} 
\[
    \fl
    \Phi(\phi(x),\xi)
    \approx {\cal N}\exp\left[-\frac12 \int d^3x\, d^3y \frac{\sqrt{\gamma(\bf x,\xi)}}{N(\bf x,\xi)} 
      \sum_i \frac{dK_i(\bf x,\xi)}{d\xi}K^{-1}_i(\bf y,\xi) \phi(\bf x,\xi) \phi(\bf y,\xi)\right]\,, 
\]
where the Euclidean time coordinate $\xi$ specifies the time slicing, and 
$N$ and $\gamma$ are the lapse function and the determinant of the 
induced metric, respectively. 
If we expand $\phi({\bf x},\xi)$ as 
\[
    \phi({\bf x},\xi)=\sum_j \phi_j(\xi)K_j({\bf x},\xi)\,, 
\]
we have 
\begin{equation}
    \fl
    \Phi(\phi(x),\xi)
    \approx {\cal N}\exp\left[-\frac12 \int d^3x\, 
      \frac{\sqrt{\gamma(\bf x,\xi)}}{N(\bf x,\xi)} 
      \sum_{ij} \frac{dK_i({\bf x},\xi)}{d\xi}K_j({\bf x},\xi) \phi_i(\xi) \phi_j(\xi)\right]\,.  
      \label{eq:eq13}
\end{equation}
Hence, if 
either $K_i({\bf x},\xi)$ or $dK_i({\bf x},\xi)/d\xi$ flips 
the sign for all ${\bf x}$ at some $\xi$, 
the wave function definitely becomes unnormalizable in the direction of 
$\phi_i$, at least, within the Gaussian approximation.

We will later apply this general idea to a specific problem of true vacuum decay/upward tunneling. 
Before discussing this more exotic case, we will briefly review a much more popular false vacuum decay described by an  $\mathrm{O} (4)$-symmetric instanton in the next section.

\section{\label{sec:rev}False vacuum decay with gravity}

As is known, there is a consistent description of the false vacuum decay to a true vacuum decay coupling with gravity. 
In this case the dominant contribution to the tunneling rate is coming from an $\mathrm{O} (4)$ symmetric instanton, and a clear WKB picture of the process can be given. 
The initial false vacuum is described by pure de~Sitter space with a positive cosmological constant, while the final true vacuum appears inside a bubble formed in the false vacuum sea. 
The $\mathrm{O} (4)$ symmetric instanton consistently interpolates between these two configurations. We report the essential points of the discussion in section~2 of Ref.~\cite{bib:PrThP1992..88...503S} to fix the notation and introduce some important equations. 
The Euclidean action describing the process
is
\[
	S _{\mathrm{E}}
	=
	\int d ^{4} x \sqrt{g}
	\left[
		- \frac{1}{2 \kappa} R
		+
		\frac{1}{2} g ^{\mu \nu} \partial _{\mu} \phi \partial _{\nu} \phi + V ( \phi )
	\right]
	\,.
\]
The absolute minimum of the potential $V ( \phi )$ is denoted by $\phi _{\mathrm{T}}$, while another local minimum is $\phi _{\mathrm{F}}$.
After choosing coordinates $(\tau, \chi , \theta , \phi)$, where $\chi$, $\theta$, and $\phi$ are the usual coordinates on a $3$-sphere, ${\mathbb{S}} ^{3}$, we consider an Euclidean metric that in these coordinates takes the form
\begin{equation}
	ds^2
	=
	N ^{2} d\tau^2+ a ^{2}\left(d\chi^2+ \sin ^{2}\!\chi \left(d\theta^2+ \sin^{2}\!\theta d\phi^2\right)\right)
	 \,,
\label{eq:inimetten}
\end{equation}
where $N$ and $a$, as well as the scalar field $\phi$, are functions of the Euclidean time coordinate $\tau$ only. 
Under these assumptions, the relevant equations of motion for the system are
\begin{eqnarray}
	\ddot{\phi} + \left( 3 \frac{\dot{a}}{a} - \frac{\dot{N}}{N} \right) \dot{\phi}
	=
	N ^{2} \frac{d V}{d \phi}\,,
	\nonumber \\
	\dot{a} ^{2} - \frac{\kappa}{6} a ^{2} \dot{\phi} ^{2}
	=
	N ^{2} \left( 1 - \frac{\kappa}{3} a ^{2} V ( \phi ) \right)
	\,,
	\nonumber
\end{eqnarray}
where an overdot denotes a differentiation with respect to $\tau$, and the second equation is nothing but the Hamiltonian constraint. Euclidean de~Sitter space is realized by the solution $\phi ( x ) \equiv \phi _{\mathrm{F}}$, with $V (\phi _{\mathrm{F}}) > 0$. For this constant value of the scalar field that locally minimizes $V (\phi)$, the first of the above two equations is clearly satisfied, while the second now reads
\[
	\left( \frac{1}{N}\frac{d a}{d \tau} \right) ^{2}
	=
	1 - a^2 \hat H _{\alpha} ^{2} 
	\,, \quad
	  \hat H _{\alpha} ^{2} := \frac{\kappa V ( \phi _{\alpha} )}{3}
	\,, 
\]
where $\alpha =$ ``T'' or ``F''. 
The scale factor is, then, determined as
\[
	a (\tau) = a_{\alpha}(\tau):=\hat H^{-1}_{\alpha} \sin \left( \hat H_{\alpha}\int ^{\tau} N(\tau') d \tau' \right)\,.
\]
For simplicity, we make the gauge choice $N \equiv 1$. 
There is a bounded interval of values of $\tau$ that are meaningful for the problem, say, $[\tau _{\mathrm{i}} , \tau _{\mathrm{f}}]$, and the scale factor vanishes at the boundaries. 
The appropriate boundary conditions required by regularity are
\[
	\dot{\phi} ( \tau _{\mathrm{i}} )
	=
	\dot{\phi} ( \tau _{\mathrm{f}} )
	=
	0
	\ .
\]
At least two solutions of this kind can be considered, which are known in the literature as the Coleman--De~Luccia (CDL) solution~\cite{bib:PhReD1980..21..3305L}, and the Hawking--Moss solution~\cite{bib:Plett1982110B....35M}. Here, we are interested in the CDL type, and soon we will make the additional assumption to be in the \emph{thin-wall approximation}. The existence of this solution is not guaranteed in contrast to the Hawking-Moss one, but it dominates the tunneling as long as the peak of the potential is sufficiently sharp~\cite{Jensen:1983ac,Jensen:1988zx,Samuel:1991dy}. Additionally, other solutions with less symmetry than $\mathrm{O} (4)$, e.g. $\mathrm{O} (3,1)$, have been considered (see, {\it e.g.}, the seminal works~\cite{bib:PhReD1987..35..1747G,bib:PhReD1987..36..2919T} among the vast literature on this subject).

\begin{figure}[tb!]
	\begin{center}
		\includegraphics{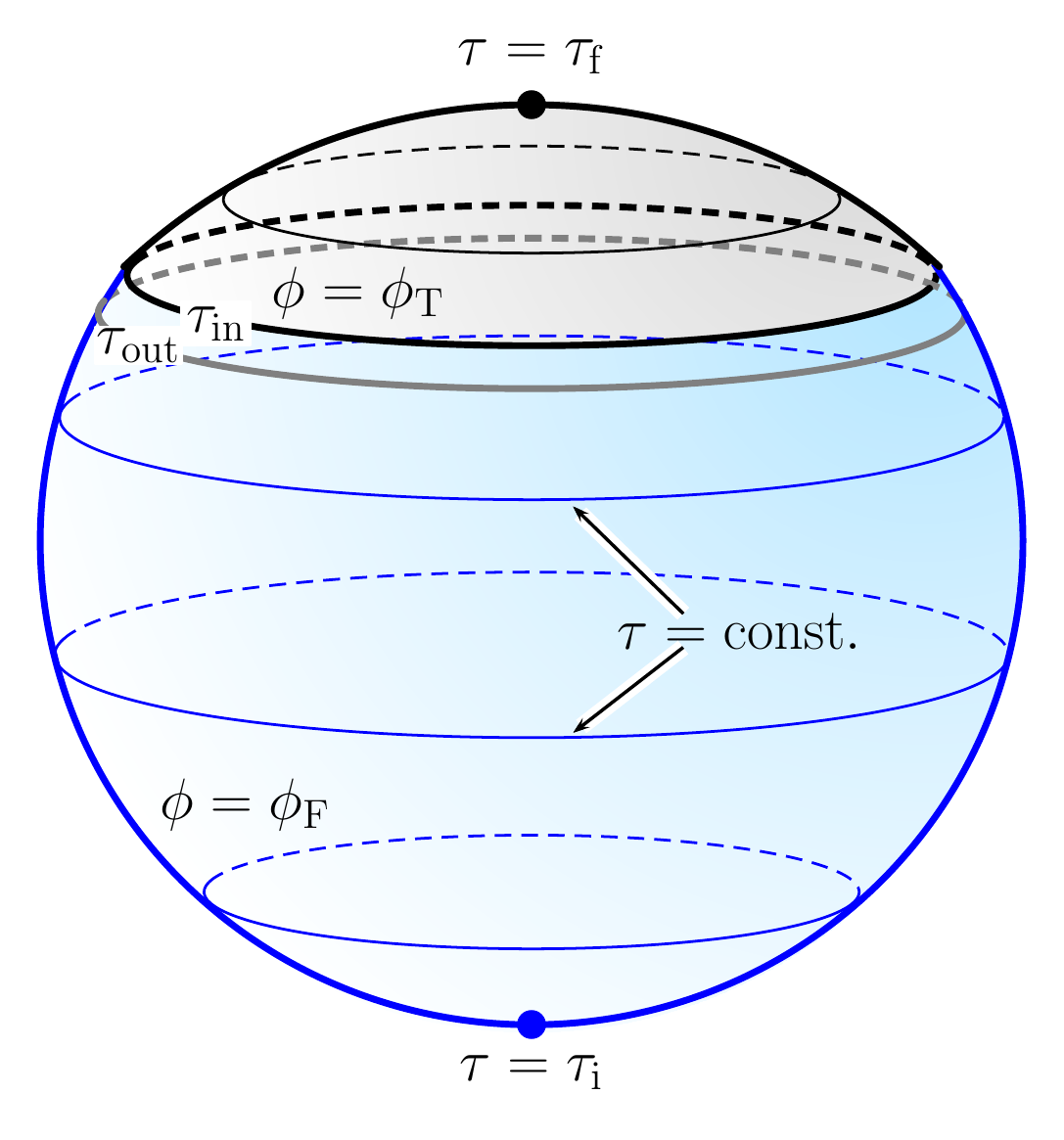}
		\caption{\label{fig:O_4thiwal}\small{}A visualization of an $\mathrm{O} (4)$ geometry. True vacuum is at the top, and we assume $V (\phi _{\mathrm{T}}) = 0$; it covers the region between $\tau = 0$ and $\tau = \tau _{\mathrm{in}}$. False vacuum is to the bottom of the $\tau _{\mathrm{out}}$ circle. The \emph{thin wall approximation} considers that the transition region between $\tau = \tau _{\mathrm{in}}$ and $\tau = \tau _{\mathrm{out}}$ is small when compared with the other scales in the problem. Picture reproduced from~\cite{bib:PrThP1992..88...503S}. 
   }
	\end{center}
\end{figure}

In the thin-wall limit, we assume that the system for most of the evolution sits at either of the two minima of the potential, and that the transition between them takes place over a very narrow interval in $\tau$. This is shown in figure~\ref{fig:O_4thiwal}, where we see a \emph{true} vacuum bubble (the spherical cap at the top having a larger curvature radius, with $V ( \phi _{\mathrm{T}} ) = 0$ and bounded by $\tau _{\mathrm{in}}$) surrounded by false vacuum (the spherical region to the bottom of $\tau _{\mathrm{out}}$). The transition between these two regions takes place in the narrow interval between $\tau _{\mathrm{in}}$ and $\tau _{\mathrm{out}}$. 

To construct a WKB wave function corresponding to the false vacuum decay starting with the instanton shown in figure~\ref{fig:O_4thiwal}, we need to find an appropriate slicing such that:
\begin{enumerate}
	\item the initial slice, which we call $\Sigma _{\mathrm{i}}$ following \cite{bib:PrThP1992..88...503S}, describes a field configuration in the false vacuum and the spatial geometry before the tunneling;
	\item the final slice, $\Sigma _{\mathrm{f}}$, contains a bubble of true vacuum surrounded by a false vacuum region. 
\end{enumerate}	
Both, the initial and final slices, can be connected by analytic continuation to the Lorentzian configurations and geometries that describe (i) the classical state in the false vacuum before the tunneling, and (ii) the evolving true vacuum bubble in the false vacuum sea after the tunneling, respectively.

\begin{figure}[tb!]
	\begin{center}
		\includegraphics{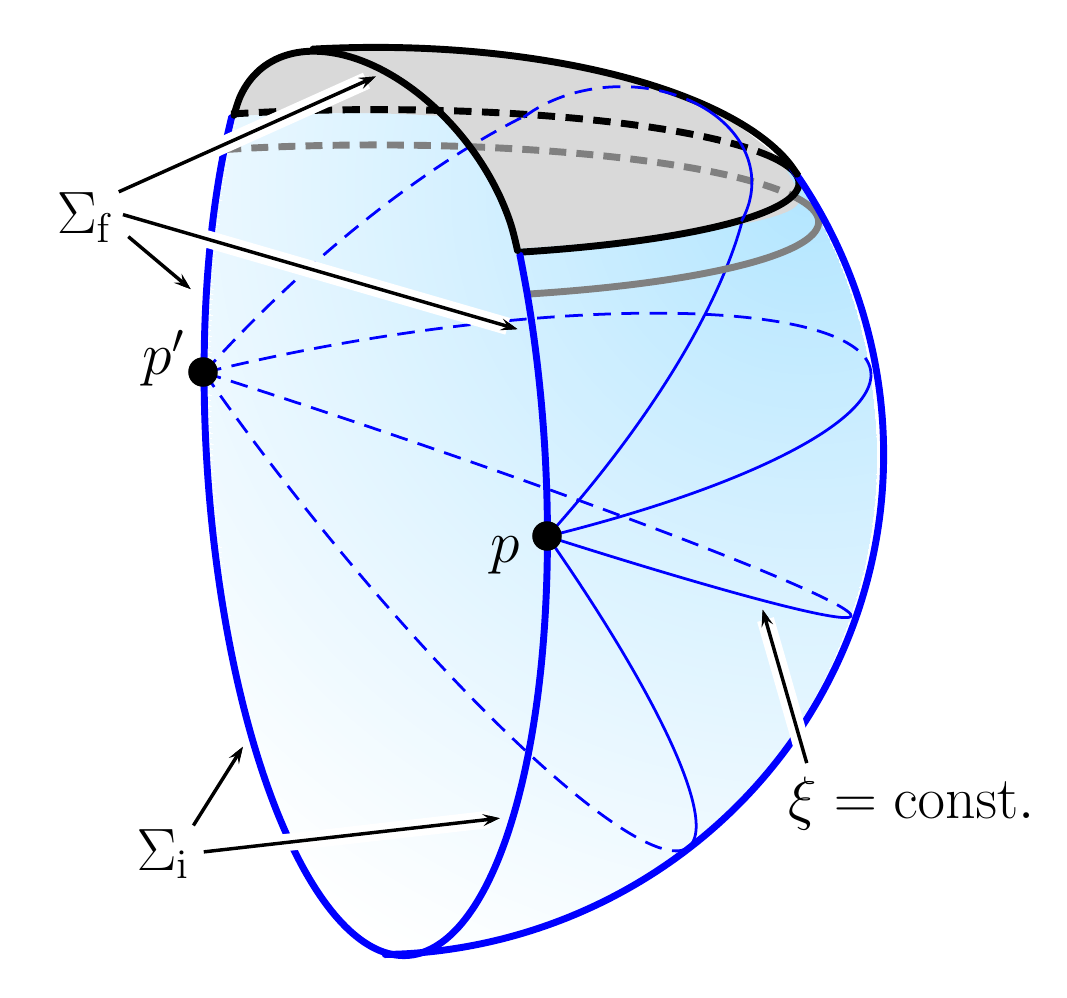}
		\caption{\label{fig:O_4thiwalins}\small{}A representation of the thin wall instanton associated to a true vacuum bubble nucleation inside false vacuum. The initial and final slices are denoted as $\Sigma _{\mathrm{i}}$ and $\Sigma _{\mathrm{f}}$, respectively. The picture shows a foliation according to Euclidean de~Sitter time, $\tau$. Additional details, e.g., about the transformation from the coordinates used in~(\ref{eq:inimetten}) to de~Sitter coordinates, can be found in~\cite{bib:PrThP1992..88...503S}, from where the image is reproduced. 
}
	\end{center}
\end{figure}

The continuous foliation, which uses a half of the $\mathrm{O} (4)$ symmetric instanton of Fig.~\ref{fig:O_4thiwal}, is shown in Fig.~\ref{fig:O_4thiwalins}. In this picture $\Sigma_{\rm i}$ is a time constant slice of a single patch in the static chart of Euclidean de~Sitter space. The points $p$ and $p'$ are isolated in this 2-dimensional figure, but 
they are a connected 2-surface with the de~Sitter horizon radius in the case of four dimensional spacetime. 
To establish a WKB interpretation, the initial state must be stationary. Therefore, we need to consider the 
constant time surface of the static chart. Basically, the boundary condition is kept unchanged during the 
Euclidean time evolution, \textit{i.e.}, the Dirichlet boundary condition is imposed. After the tunneling, and a bubble is formed completely within a static patch. The foliation is shown as $\tau=$constant surfaces in Fig.~\ref{fig:O_4thiwalins} and gives a continuous deformation of the field configuration and of the spatial geometry, which defines the dominant escape path.  

When the instanton can be constructed, and when it can be continuously connected with the before/after tunneling Lorentzian evolution by analytic continuation as in the CDL case, it is easy to justify the estimate of the transition probability associated to the dominant instanton contribution by considering the WKB wave function along the dominant escape path. 
Details can be found in the literature~\cite{bib:PrThP1992..88...503S}. Here, however, we should point out a property of this ordinary instanton that will be crucial for our discussion later. 
Indeed, in this case, the lapse function, which specifies the time slicing in moving from one slice to the next, can be chosen to be nowhere negative.
As a result, this very same choice \emph{covers every point of the Euclidean manifold exactly once}. 
At first sight, this property might be given for granted, if one could na\"{\i}vely think that it is always possible to find such a foliation. In fact, this is not always the case~\cite{bib:NuPhy1990B339...417G,bib:JETPh2015.120...460T} (see also~\cite{bib:PhReD2005..72103525J} for an additional discussion of problems related to the tunneling configurations).  This is exactly what we are going to consider in detail in what follows, by discussing if it is justified to apply the WKB argument to calculate the tunneling rate also in cases where no foliation with everywhere positive lapse and single covering of the instanton manifold can be found. We will start in the next section, by first discussing a simplified model, in which the WKB approximation does not provide a consistent description.

\section{\label{sec:wkbbre}Breakdown of the WKB approximation in a simple model}

\begin{figure}[tb!]
	\begin{center}
		\includegraphics{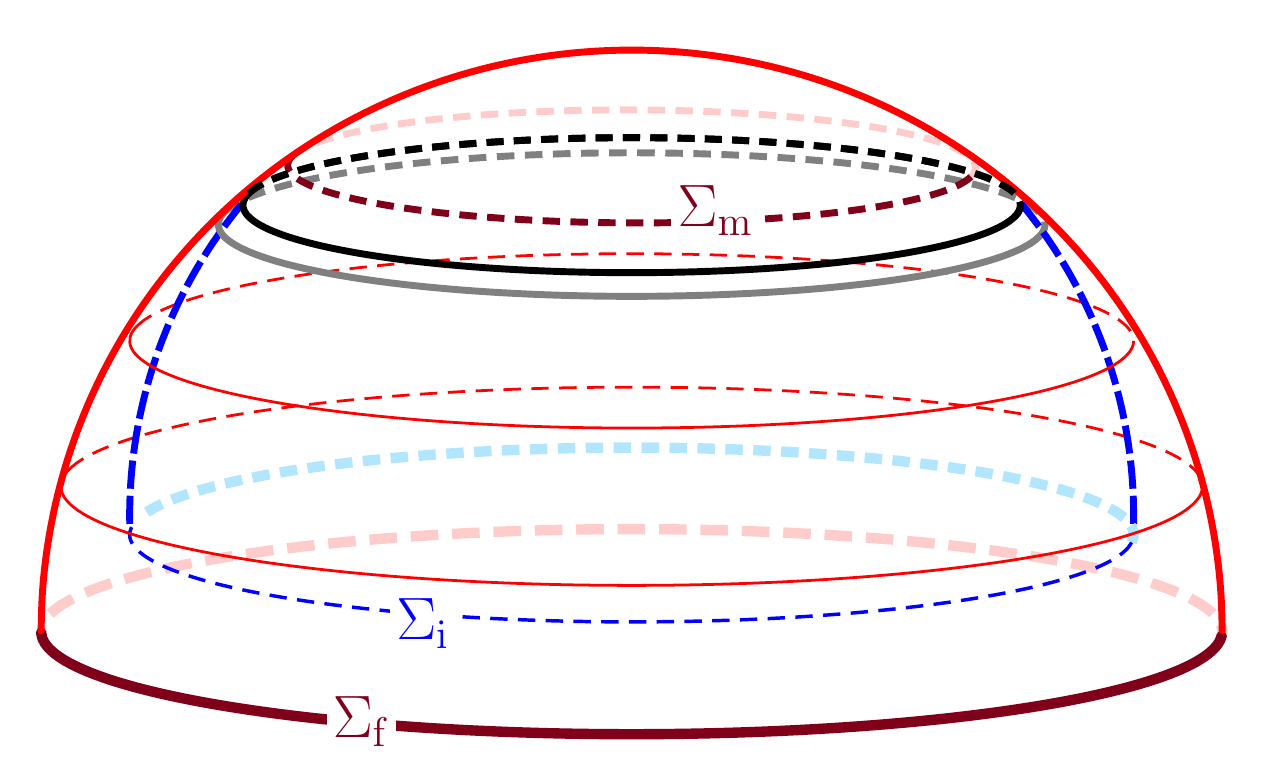}
		\caption{\label{fig:O_4truvacdec}\small{}
  The Euclidean solution representing 
  a simplifed upward tunneling process.   
 Parts of the red (bigger) and blue (smaller) spheres 
 describe regions in the true and false vacua, respectively. 
When the wall is present, the upper red spherical cap above $\tau _{\rm in}$ is connected to the di Sitter space with a higher vacuum energy density, i.e. the blue part, while 
the upper spherical cap is connected to the red side 
when the wall is absent. 
Two possibilities are simultaneously are drawn in this spacetime diagram. 
}	\end{center}
\end{figure}

Here, we consider a simplified model of tunneling. Namely, the transition between 
two de~Sitter spaces at different minima of the potential.  
The Euclidean solution to consider is the one shown in Fig.~\ref{fig:O_4truvacdec}.  
Here, the red and blue parts describe regions of de~Sitter space in the true and false vacua, respectively. 
When the wall is present, the upper red spherical cap above $\tau _{\rm in}$ must be thought as connected to the higher vacuum energy density de~Sitter space, i.e. the blue part, while 
the red part remains to be red in the absence of the wall. 
Hence, two possibilities are simultaneously drawn in this spacetime diagram. 
The initial slice for the tunneling that we consider here is a great circle 
on the side without the wall, 
while the final slice is on the side with the wall. 
It is immediately transparent that 
we cannot find a dominant escape path in a single manifold. 
Nevertheless, we can find a path of approximate Euclidean time evolution that 
connects the initial and final slices. 
For this purpose, we need to pass through an intermediate time slice, $\Sigma _{\mathrm{m}}$, 
fully contained in the upper cap region. 
Starting with the side without the wall, we switch to the solution with the wall at this intermediate time slice. Then, we can find a remaining part of the path that connects to 
the final slice. 

For definiteness, we consider the horizontal foliation of this spacetime such that the lapse function changes sign in the upper cap region above the wall. In terms of the corresponding scale factors, $a _{\mathrm{T}}$ and $a _{\mathrm{F}}$, that describe the evolution of the background geometries, we have the usual relations
\begin{equation*}
    H_\alpha:=\left(\frac{\dot{a}_{\alpha}}{a_\alpha}\right)=-\sqrt{\frac1{a^2_{\alpha}} - \hat{H}_{\alpha}^2} \,, 
\label{eq:Hubfundef}
\end{equation*}
with $\alpha=$ ``T'' in the larger sphere, while $\alpha=$ ``F'' in the smaller sphere, 
and recall that $\hat H_{\rm F}$ and $\hat H_{\rm T}$ satisfy 
$\hat H_{\rm F} > \hat H_{\rm T}$. Here, an overdot represents the $\tau$-derivative, and 
this time coordinate, $\tau$, is related to the parameter of the foliation, $\xi$, by 
$d\tau=N d\xi$. Then, along the dominant escape path, $\xi$ monotonically increases, while 
$\tau$ takes the maximum value where $N$ flips the sign at the intermediate time slice $\Sigma _{\mathrm{m}}$. The  sign in front of the square root appearing in the definition of $H _{\alpha}$ is negative, as we choose $\tau$ to increase with decreasing scale factor.

We now concentrate on the high momentum limit of the field equations for a massless scalar field, where we can safely neglect the gravitational interaction. On this background the mode function $K_L(\tau)$ decomposed using the spherical harmonics on a unit 3-sphere should satisfy
\[
	\ddot{K}_L + 3 H_\alpha \dot{K}_L - \frac{L(L+2)}{a_{\alpha}^2}K_L = 0 \,, \quad \alpha = \mathrm{T} , \mathrm{F}\,,
\]
where $L(L+2)~(\gg 1)$ is the negative of the eigenvalue of the Laplacian operator on the unit 3-sphere and the field is assumed to be massless. 
Let us focus on a small region in which the scale factor, $a$, and $H$ are approximately constant 
with the values that they take at the wall radius, {\it i.e.,} $a _{\mathrm{T}/\mathrm{F}}\approx a_{\rm w}$ and 
$H _{\alpha}\approx H_{{\rm w}\alpha} = - (a _{\rm w} ^{-2} - \hat{H} _{\alpha} ^{2} ) ^{1/2}$. 
Under this approximation, substituting the standard ansatz $K_L
\sim \exp (\lambda \tau)$ results in the characteristic equation
\[
	\lambda ^{2} + 3 H_{{\rm w}\alpha} \lambda - p ^{2} = 0
	\,,
\]
where $p^2=L(L+2)/a_{\rm w}^2$. 
The two roots can be approximated for large $p$ as follows
\begin{eqnarray}
	\lambda _{\alpha\pm}
	& = &
	\frac{-3 H_{{\rm w}\alpha} \pm \sqrt{9 H_{{\rm w}\alpha} ^{2} + 4 p ^{2}}}{2} \cr
    &\approx &\pm p - \frac{3}{2} H_{{\rm w}\alpha}
	\,.
\end{eqnarray}
Then, the growing mode describing the first part of the process before reaching $\Sigma _{\mathrm{m}}$ is
\[
	\phi_{\rm T} \sim \exp{(\lambda _{{\rm T}+} \tau)}\,.
\]
For simplicity, we set the origin of $\tau$ at the position of the wall, 
where the solution has to be joined smoothly to the one on the false vacuum side. 
The solution after the junction is given, in general, by 
\[
	\phi_{\rm F} \sim c_1 \exp{( \lambda _{{\rm F}+} \tau )} 
         + c_2 \exp{( \lambda _{{\rm F}-} \tau )}
	\,,
\]
where $c_1$ and $c_2$ are real constants. 
Continuity up to the first derivative of the scalar field results in the junction conditions
\begin{eqnarray}
	1 = c_1 + c_2\,,
	\nonumber \\*
	\lambda_{{\rm T}+} = c_1 \lambda_{{\rm F}+} + c_2 \lambda_{{\rm F}-}
	\nonumber
	\,,
\end{eqnarray}
which determines $c_2$ as
\[
	c_2 \simeq \frac{3}{4 p} ( H_{\rm wT} - H_{\rm wF} )
	\ .
\]
From our discussion above, we see that $H_{\rm wT}$ and $H_{\rm wF}$ are both negative, 
and $H_{\rm wF} > H_{\rm wT}$. Hence, in the large momentum limit, we have $0< -c_2 \ll 1$, and thus $c_1 \approx 1$. 
By considering our approximate solution, one can easily 
find a zero of $\phi_{\rm F}$ at 
\[
	\tau= \tau _{\mathrm{c}}
	=
	\frac{1}{\lambda _{{\rm F}+} - \lambda _{{\rm F}-}} 
        \log \left( -\frac{c_2}{c_1} \right)
	\approx \frac{1}{2p} \log \left(\frac{3}{4 p} ( H_{\rm wF} - H_{\rm wT} )\right)<0
	\ .
\]
This expression tells us  
in the large $p$ limit $\tau _{c}$ is arbitrarily close to $0$, {\it i.e.,} arbitrarily close to the wall. 
Since the wall is not infinitesimally thin in reality, the above analysis does not 
apply to infinitely large $p$, but it will apply to $p$ comparable to the inverse of the 
wall thickness. Thus, $\tau_c$ should be very close to the wall, as long as the wall is thin. 
As discussed in Sec.~\ref{sec:WKB}, the sign flip of 
the mode, or its derivative, implies the breakdown of the WKB approximation. 

One may wonder if it is appropriate to choose the growing solution for $K_L$ at the beginning. 
If we are allowed to start initially with the decaying solution, the above mentioned pathology might be avoided.  
However, this idea does not work. 
If we wish to use the decaying solution, we need to flip the sign in the exponent in Eq.~(\ref{eq:eq13}), where 
we assume that $N$ is positive to avoid confusion. 
To maintain that the WKB wave function satisfies the required wave equation, we need to 
flip the sign in the exponent in Eq.~(\ref{eq:Psi0}), too. This implies that the amplitude of the 
wave function under the potential barrier becomes larger than that for the initial configuration, which 
is definitely contradictory because the probability of taking a configuration under the barrier becomes 
larger than the probability for the initial configuration. 
Therefore, this possibility is not physically acceptable. 

The current analysis is based on a particular model. However, 
as we are focusing on high momentum modes, this breakdown of WKB approximation is 
expected to be a universal feature near the vanishing lapse region, 
independently of the details of the background instanton solution. We are going to elaborate on this point in the following section.

\section{Other possibly pathological tunneling processes\label{sec:othpat}}

\begin{figure}[tb!]
	\begin{center}
		\includegraphics{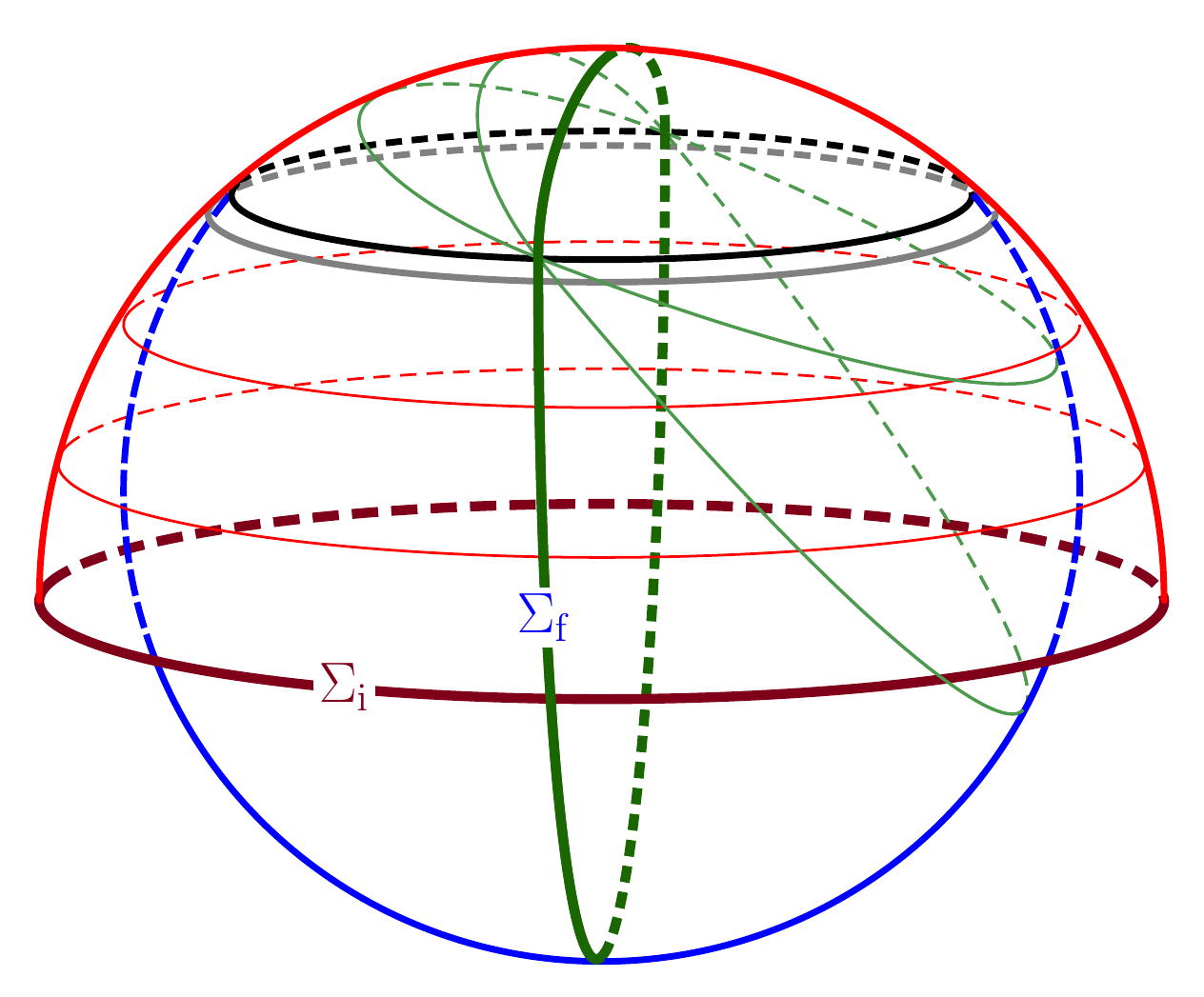}
		\caption{\label{fig:O_4truvacdec_02}\small{}Schematic picture that describes up-ward tunneling. Two solutions are connected exactly in the same way as in the case of Fig.~\ref{fig:O_4truvacdec}. Only the choice of the time slicing is different. The final time slice contains a true vacuum bubble inside a false vacuum sea.}
	\end{center}
\end{figure}

\subsection{Upward tunneling}
Here, we discuss the \emph{upward tunneling}~\cite{bib:PhReD1987..36..1088W}, also called \emph{true vacuum decay}, {\it i.e.,} the transition from a lower energy vacuum to a higher energy vacuum. 
The initial state is given by de~Sitter spacetime at the true vacuum, while 
the end state is the same as the case of the false vacuum decay. This final 
configuration can be interpreted as such that a false vacuum nucleates in a 
true vacuum sea. In contrast to the case of false vacuum decay, the nucleated 
bubble does not fit within a single static chart of the initial de~Sitter spacetime. 
Some discussion about the true vacuum decay can be found in~\cite{bib:PhReD1987..36..1088W}, for the case in which the true vacuum is a de~Sitter space with a non-vanishing vacuum energy. 
We will show that Euclidean solutions that realize a true vacuum decay must include points, at which the lapse function becomes negative. 

In the case of the upward tunneling the construction of the dominant escape path is quite subtle. 
If we consider the same instanton shown in Fig.~\ref{fig:O_4thiwalins},
we might be able to identify the union of $\Sigma_{\mathrm{f}}$ and $\Sigma_{\mathrm{i}}$ as the final slice. 
However, in this manifold we cannot find the initial slice, 
which must be a slice containing only true vacuum. 

To find a dominant escape path, 
we need to consider an Euclidean evolution as the one in Fig.~\ref{fig:O_4truvacdec}, discussed in the 
preceding section. 
Again, we can find the requested initial slice, $\Sigma _{\mathrm{i}}$, on the larger sphere in Fig.~\ref{fig:O_4truvacdec}. 
Once we evolve into the top cap region, we switch to the configuration with the wall. 
Hence, now the lower part is replaced by the smaller sphere. Then, one can find a slice, $\Sigma _{\mathrm{f}}$, 
corresponding to the final state of tunneling, by considering the foliation shown in Fig.~\ref{fig:O_4truvacdec_02} (in the figure, we use green for the slices that result in the final configuration containing a false vacuum bubble immersed in the true vacuum). 
In this way, it is possible to construct a dominant escape path that describes the upward tunneling process, but vanishing of the lapse function is required in the blue region as in the case of 
the simple example discussed in the preceding section. Therefore, we cannot avoid that 
the WKB wave function for the quantum fluctuation becomes unnormalizable, as before. 
This is in contrast with the false vacuum decay case in the thin wall approximation, where we can always avoid the vanishing of the lapse.

\subsection{False vacuum decay catalyzed by a black hole}

\begin{figure}[tbh]
    \centering
    \includegraphics{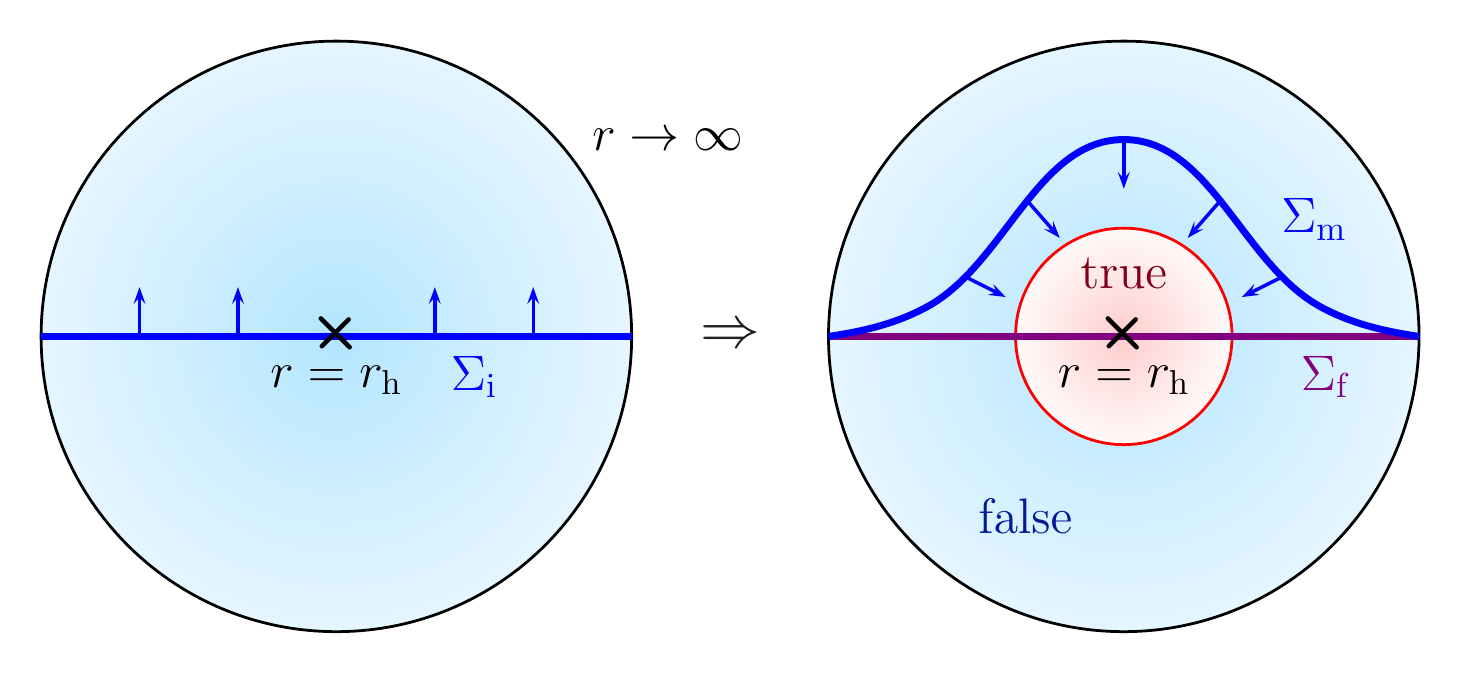}
    \caption{The Euclidean solution that describes a bubble nucleation catalyzed by a black hole. Here the radial direction is corresponding to the ordinary radial direction while the angular direction is corresponding to the Euclidean time. The initial surface $\Sigma_{\rm i}$ is deformed in the original Euclidean Schwarzschild spacetime in the left panel.  Then, we switch to the configuration containing a nucleated bubble in the right panel, and the time slice is deformed back reaching the final surface $\Sigma_{\rm f}$. }
    \label{fig:bhc}
\end{figure}

In the next example we start with a black hole solution in a false vacuum sea. 
For simplicity, setting the vacuum energy in the false vacuum to zero, we 
assume the initial configuration to be a simple Schwarzschild spacetime. 
Its analytic continuation to the Euclidean solution through a maximal surface, 
which is the so-called Euclidean Schwarzschild solution, gives 
a cigar-like geometry when we suppress the angular directions. 
Here, we use a conformal transformation to map this cigar-like configuration to a disk, as shown in Fig.~\ref{fig:bhc}.

From this initial state, we consider the process of nucleation of a true vacuum bubble. 
After a true vacuum bubble with negative vacuum energy density is formed, the configuration is given by a junction of a Schwarzschild spacetime and a Schwarzschild-anti de~Sitter (AdS) spacetime, separated by a domain wall. 
For simplicity, we consider the critical case, in which 
the analytic continuation to the Euclidean solution is obtained by 
a small disk with a constant radius around the center of the Euclidean Schwarzschild disk in Fig.~\ref{fig:bhc}, 
and connected to the Euclidean Schwarzschild-AdS disk with the same circumferential radius. 
As we cannot find the maximal slice corresponding to the initial configuration in this Euclidean solution, 
we need to look for a dominant escape path as in the examples already mentioned above. 
Starting with a maximal surface in the Euclidean Schwarzschild solution, we evolve the time slice
to be able to avoid the disk region, which is to be replaced with the Euclidean Schwarzschild-AdS disk. 
Then, we switch to the configuration with a domain wall. 
After the switch, introducing a negative lapse, we can come
back to another maximal surface, which contains a true vacuum
bubble. 
Also in this example it seems necessary to have a flip of the sign of the lapse function along the dominant escape path. Hence, we expect that a similar pathology in the construction of the
WKB wave function is inevitable also for this process, at least in the low temperature regime, where instanton driven transitions should dominate over thermally assisted ones~\cite{bib:Briaud:2022few,bib:Gomberoff:2003zh}. 

\subsection{FGG case}

We now discuss another example, in which the same features that we considered above are present. This is the tunneling process originally considered by Farhi, Guth and Guven in~\cite{bib:NuPhy1990B339...417G}. The model in~\cite{bib:NuPhy1990B339...417G} considers a given vacuum phase, in which there is a bubble of a different vacuum phase isolated by a wall. According to the classical dynamics, there are two branches of the wall motion; one is the bounded one, having a maximum radius, and the other is the bouncing one, having a minimum radius. 
When the bubble wall is initially bound to recollapse in the classical dynamics, quantum effects could allow it to overcome the potential barrier that separates two different branches, and to be transferred to a forever expanding configuration. The non-zero probability associated to this process, {\it i.e.}, the tunneling under the potential barrier, can be estimated at the lowest order in the WKB approximation by constructing an appropriate instanton and calculating the associated classical action, as we did above. 
We consider the case in which the maximal extension of the initial 
`exterior' spacetime, which means the part outside the bubble, is given by the Schwarzschild solution, 
which has two causally disconnected asymptotic regions. 
In this case there are two qualitatively very different types of instantons: (i) the one in which the bubble wall just becomes `slightly bigger' remaining on the same outside region where it originally was; and (ii) the one in which it actually tunnels 
to the other outside region. Note that, in the initial configuration (before the tunneling) no black hole is present. However, in the latter case a black hole arises in the final configuration after the tunneling, 
which is why the process is also called, tunneling with black hole/wormhole creation. 
We are here interested in this second process.

It was early realized~\cite{bib:NuPhy1990B339...417G} that some conceptual and technical complications affect this second process. In particular, (a)~it seemed unclear how to consistently construct an Euclidean geometry interpolating between the two classical configurations. Moreover, (b)~the calculation of the classical action seemed plagued by ambiguities~\cite{bib:PhReD2006..73123529J}. Finally, (c)~the validity of the WKB approximation is not clear in this setup. The first and second problems have been raised in the original paper by Farhi, Guth, and Guven, and can be addressed by carefully considering all the boundary contributions to the action in a smooth foliation of the instanton~\cite{bib:JETPh2015.120...460T}. Indeed, it is important to 
find an instanton that realizes a smooth foliation, and then to take into account all the necessary boundary terms to compute the magnitude of the action. Complication (c) was not explicitly considered in this specific scenario, although it is not new in the context of vacuum decay~\cite{bib:Plett1982110B....35M,bib:PrThP1992..88...503S}, and it is what we are concentrating on here.

\begin{figure}[tb!]
	\begin{center}
		\includegraphics{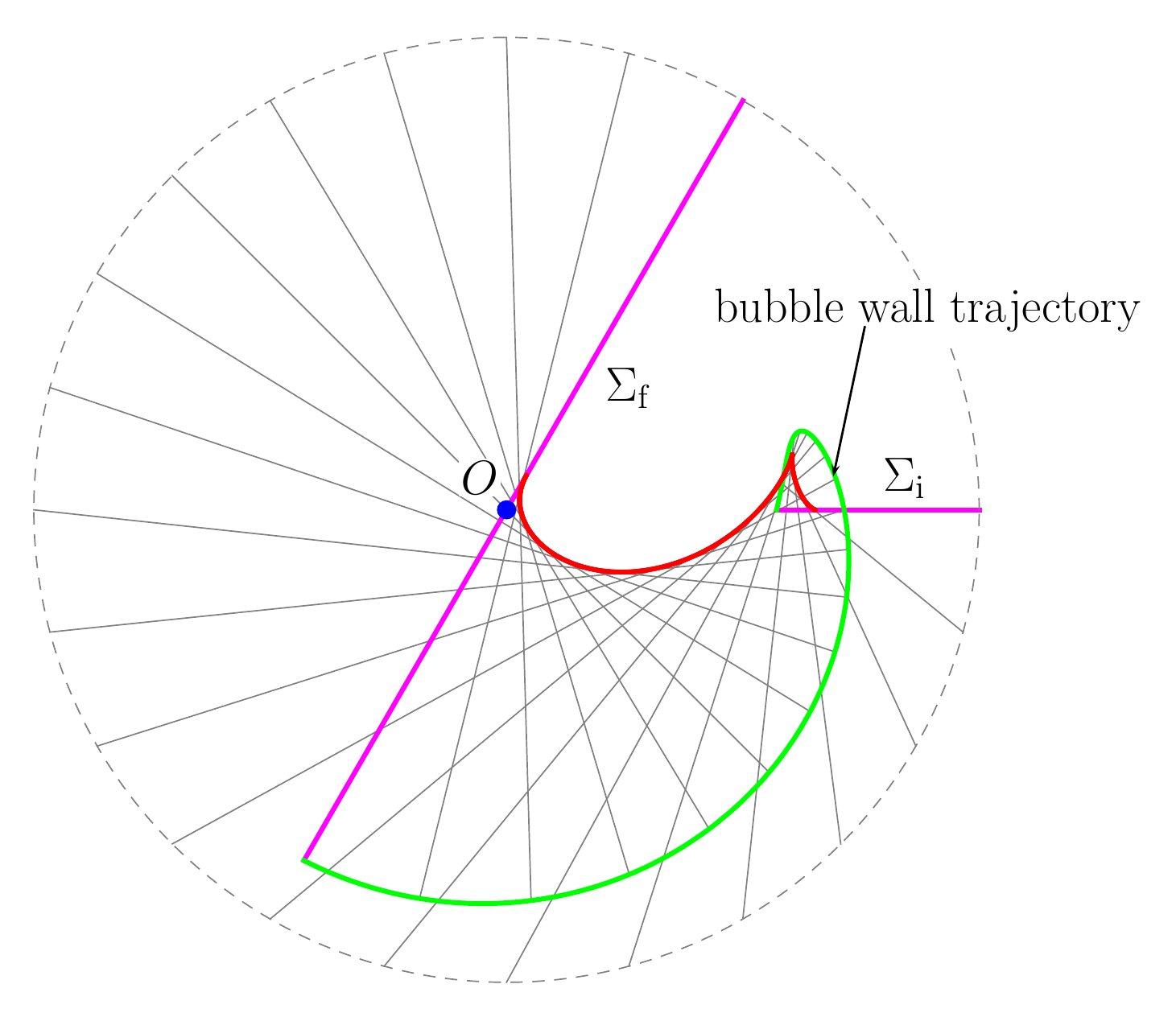}
		\caption{\label{fig:FGGmulcovvanlap}\small{}We show an arbitrary smooth foliation (gray lines) connecting the exterior parts of the initial slice, $\Sigma _{\mathrm{i}}$ to the exterior part of the final slice $\Sigma _{\mathrm{f}}$. As defined in the text, and following reference~\protect\cite{bib:NuPhy1990B339...417G}, a covering number must be associated to the instanton, which takes into account how many times, and in which direction (clockwise or anticlockwise), the foliation sweeps each point of the instanton. In this foliation, the necessity of the covering number is made evident by the fact that the slices cross each other. This instanton mediates vacuum decay with wormhole production, as the initial slice does not contain the bifurcation point, $O$, but the final slice does.}
	\end{center}
\end{figure}

In particular, let us consider the Euclidean trajectory of the bubble wall, {\it i.e.}, the evolution of the surface of separation between the two vacuum phases, in the case of tunneling with wormhole production. Figure~\ref{fig:FGGmulcovvanlap} describes the Euclidean Schwarzschild spacetime, suppressing the two dimensions of the maximally symmetric sphere. 
The center of this picture, $O$, corresponds to the event horizon, or the bifurcation two sphere, and the boundary circle is the asymptotic infinity. 
The green thick solid curve is the trajectory of the wall, and the surfaces $\Sigma_{\rm i}$ and $\Sigma_{\rm f}$ are the exterior parts of the initial and final slices, respectively. 

The instanton is, then, characterized by the following properties:
\begin{enumerate}
    \item the initial and final slices are both a part of a straight line passing through the center $O$, so that they are maximal surfaces;
    \item the bubble wall trajectory is normal to the initial and final slices, since the `momentum' of the wall should vanish at the turning points;
    \item the initial slice does not contain $O$, while the final slice does, which characterizes tunneling with black hole/wormhole production;
    \item a smooth family of slices (foliation) interpolating between the initial and the final ones, and each of them extends up to infinity. (Recall that we are considering only the `outside' spacetime.)
\end{enumerate}

With reference to Fig.~\ref{fig:FGGmulcovvanlap}, let us now consider an arbitrary smooth foliation. 
Following~\cite{bib:NuPhy1990B339...417G}, we associate to each point of the instanton a \emph{covering number}, 
which counts how many times the slices of the foliation sweep the given point. 
The counting works by adding +1 (-1), to the covering number every time a smooth foliation sweeps the point counterclockwise (clockwise). 
The covering number for each point is free from the arbitrariness of the foliation, and hence it is sufficient to consider the easiest foliation. 
We immediately see that the foliation shown in Fig.~\ref{fig:FGGmulcovvanlap} attributes covering numbers other than unity to several regions. 
When the instanton has non-trivial covering numbers, the flip of sign of the lapse function takes place somewhere along any chosen foliation. 
This can be immediately shown by {\it reductio ad absurdum}. 
If there exists a foliation with positive definite lapse function, then the covering number 
must be unity everywhere on the instanton. 

The fact that the covering number of the instanton shown in Fig.~\ref{fig:FGGmulcovvanlap} is not uniformly unity tells that the appearance of negative lapse is inevitable, which may suggest the appearance of zeros 
of the mode function or its derivative, which leads to the breakdown of WKB approximation. 
However, one may think that, when a single point is swept several times, 
the mode function just reproduces the same value at every sweep. 
We discuss below that this is not the case. 

\begin{figure}[tb!]
	\begin{center}
		\includegraphics{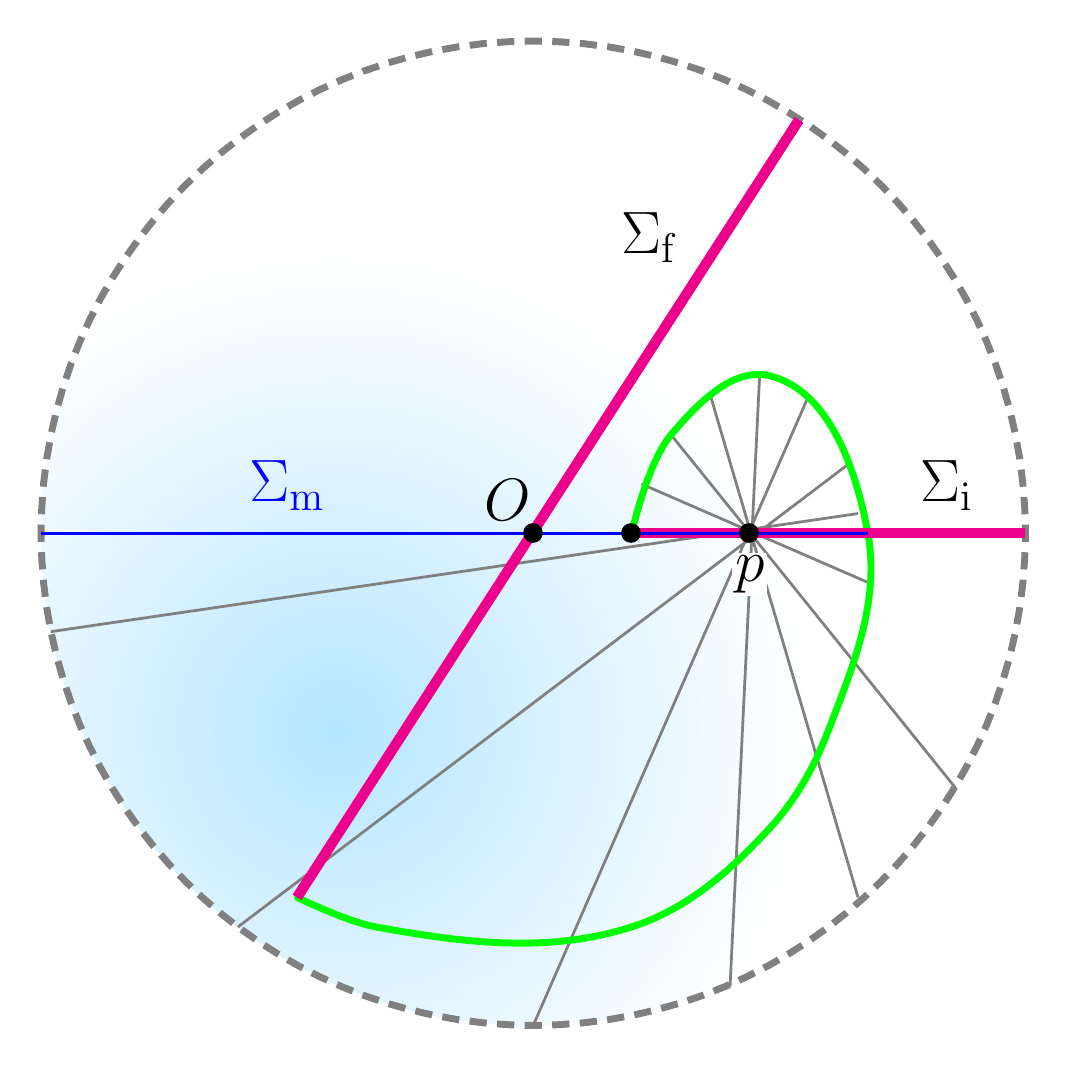}
		\caption{\label{fig:modFunWor}\small{}In this figure we choose a different foliation for the same process considered in Fig.~\protect\ref{fig:FGGmulcovvanlap}. In this foliation we can see that there is no reason why the mode function on $\Sigma _{\mathrm{m}}$ should be single valued on the segment of $\Sigma _{\mathrm{m}}$ that overlaps with $\Sigma _{\mathrm{i}}$. This inconsistency is especially clear on the right side of $p$ (see the main text).}
	\end{center}
\end{figure}

For this purpose, let us consider one particular foliation connecting 
$\Sigma_{\rm i}$ to $\Sigma_{\rm m}$ as shown in Fig.~\ref{fig:modFunWor}. 
This foliation starts with the 180$^\circ$ rotation centered at the point $p$. 
Then, the mode function on $\Sigma_{\rm m}$ would take different values from the initial ones on the surface $\Sigma_{\rm i}$. 
This would be most obvious
if we compare the values of the mode function, say, on the right side of $p$ on  
these two different time slices. 
There is no reason why they should be identical. 
In fact, roughly speaking, the mode function should increase in the clockwise direction on this segment when we consider the configuration after the rotation centered at the point $p$, while it should increase in the anti-clockwise direction in the other case. 
The non-uniqueness of mode functions is already a serious pathology, which 
would be enough to distrust the WKB wave function in this case. 

To conclude, when the change of the sign of the lapse is inevitable, 
the swept area after the sign flip of the lapse is either swept not for the first time, or connected to a different manifold. 
In both cases, concisely speaking, the background instanton solution itself is not given by a simple manifold. 
The analysis given in this paper shows, rather generically, that the WKB wave function becomes pathological in such cases.

\section{\label{sec:discon}Discussion}

In this paper we examined the construction of the WKB wave functions for various tunneling processes where 
the presence of geometrical degrees of freedom is crucial. We find that some pathological feature appears 
once we take into account the fluctuations around the dominant escape paths. 
(Here, ``paths'' is plural, since one can consider various 
foliations even if we consider the same instanton solution. From this fact, one cannot 
immediately conclude that the lowest order estimate of the tunneling rate evaluating the Euclidean 
action is wrong.)
One possible pathological feature is the 
infinite broadening of the wave function, which makes the wave function unnormalizable. Another is the 
appearance of an unavoidable time slice dependence of the quantum state. Both are related to the necessity of 
a negative lapse region, which is needed to obtain the Euclidean geometry corresponding to the dominant escape path. 
The usual false vacuum decay and the creation of the universe from nothing do not suffer from any of these pathologies. 
In many cases we are more interested in the tunneling rate than the quantum state after tunneling. 
If we do not care about the quantum state after tunneling, the pathological behavior discussed in 
this paper might not be serious. 
However, our present work would be sufficient to raise some skepticism 
on the applicability of the usual estimate of the tunneling rate based on the instanton action 
for the unconventional tunneling processes, including the true vacuum decay/upward tunneling, 
false vacuum decay catalyzed by a black hole, and the tunneling with black hole/wormhole production. 

There is an example that might remove such a skepticism, which is 
the upward tunneling in 1+1 dimensions. 
If we consider a uniform electric field, pair production of charged particles on this background 
can be thought as the tunneling to a different vacuum with a different value of the electric field strength.  
At the same time one can compute the expectation value of the electric current, which includes the virtual 
current in the process of the quantum pair production. Depending on the downward or upward 
tunneling, the direction of this virtual current is opposite, and the result is consistent 
with the estimate of the tunneling rates, in both directions, based on the instanton calculation~\cite{Frob:2014zka}. 
Note that in this case the result is obtained by the calculation of 
particle production by evolving the mode function in real time, without relying on the instanton picture. 
However, in this example the tunneling degree of freedom can be identified with 
just the motion of a single charged particle, and the other degrees of freedom seems to be completely decoupled from the tunneling degree of freedom when we treat the field corresponding to the tunneling charged particle as a free field. Hence, this example can be a special exception. 

To summarize, what we have shown here is the breakdown of the simple construction of the WKB wave function 
along dominant escape paths. While the 1+1 dimensional example analogy briefly discussed here suggests that the rate estimate based on 
instanton calculation can be still valid even if we consider upward tunneling, it is difficult to justify such an estimate without relying on a completely 
independent method, which is not easily extended to other examples. 

In the past the problem of the appearance of infinitely many negative modes was raised in Ref.~\cite{bib:PhLeB1985.161...280T}. 
This problem arises even in the conventional tunneling processes like the false vacuum decay. 
It turned out that this issue can be solved by appropriately identifying the physical degrees of freedom around the tunneling 
path~\cite{bib:PrThP1992..88...503S,Tanaka:1999pj} (See also \cite{Khvedelidze:2000cp}). In a similar way the pathological phenomena pointed out in this paper might be solved by a more sophisticated treatment. 
Here, we focused on the approach to construct the WKB wave function connecting the configurations before and after the tunneling. 
As an alternative approach, one may consider the path integral approach. 
If we na\"{\i}vely identify an instanton solution whose geometry makes 
a given manifold as the stationary point of the path integral, 
we would be able to perform the summation over the quantum fluctuations around this 
instanton to obtain the one-loop correction to the tunneling rate. In such a treatment, one would not notice the pathologies discussed in this paper,
because one would not consider whether or not the instanton solution 
is really connecting the configurations corresponding to the states before and after the tunneling. 
In our opinion, however, we would need to, at least, add some correction to the tunneling 
rate estimated using such a path integral approach, whenever the instanton solution does not contain (not even within some consistent approximation)
either the initial or the final configuration. 
Since the required correction can be significantly large or completely suppress the tunneling process, 
definitely further investigations on this issue are necessary. 

\ack{
TT is supported by JSPS KAKENHI Grant Number JP17H06358 (and also JP17H06357), \textit{A01: Testing gravity theories using gravitational waves}, as a part of the innovative research area, ``Gravitational wave physics and astronomy: Genesis'', and also by JP20K03928. SA would like to heartily thank the Theoretical Astrophysics group at the Department of Physics of Kyoto University for hospitality while working on this project.
}
\section*{References}

\bibliography{references}

\end{document}